\documentclass[fleqn,usenatbib]{mnras}

\usepackage[T1]{fontenc}

\DeclareRobustCommand{\VAN}[3]{#2}
\let\VANthebibliography\thebibliography
\def\thebibliography{\DeclareRobustCommand{\VAN}[3]{##3}\VANthebibliography}

\usepackage{graphicx}	
\usepackage{amsmath}	
\usepackage{amssymb}	
\usepackage{mathrsfs} 

\title[GWs help measure cosmic curvature]{Cosmological model-independent measurement of cosmic curvature using distance sum rule with the help of gravitational waves}

\author[Yan-Jin Wang et al.]{
Yan-Jin Wang,$^{1}$
Jing-Zhao Qi,$^{1}$
Bo Wang,$^{1}$
Jing-Fei Zhang,$^{1}$
Jing-Lei Cui$^{1}$
and Xin Zhang$^{1,2,3}$\thanks{E-mail: zhangxin@mail.neu.edu.cn}
\\
$^{1}$Department of Physics, College of Sciences, Northeastern University, Shenyang 110819, China\\
$^{2}$National Frontiers Science Center for Industrial Intelligence and Systems Optimization, Northeastern University, Shenyang 110819, China\\
$^{3}$Key Laboratory of Data Analytics and Optimization for Smart Industry (Northeastern University), Ministry of Education, China
}

\pubyear{2022}

\begin{document}
\label{firstpage}
\pagerange{\pageref{firstpage}--\pageref{lastpage}}
\maketitle

\begin{abstract}
Although the cosmic curvature has been tightly constrained in the standard cosmological model using observations of cosmic microwave background anisotropies, it is still of great importance to independently measure this key parameter using only late-universe observations in a cosmological model-independent way. The distance sum rule in strong gravitational lensing (SGL) provides such a way, provided that the three distances in the sum rule can be calibrated by other observations. In this paper, we propose that gravitational waves (GWs) can be used to provide the distance calibration in the SGL method, which can avoid the dependence on distance ladder and cover a wider redshift range. Using the simulated GW standard siren observation by the Einstein Telescope as an example, we show that this scheme is feasible and advantageous. We find that $\Delta\Omega_k\simeq 0.17$ with the current SGL data, which is slightly more precise than the case of using SN to calibrate. Furthermore, we consider the forthcoming LSST survey that is expected to observe many SGL systems, and we find that about $10^4$ SGL data could provide the precise measurement of $\Delta\Omega_k\simeq 10^{-2}$ with the help of GWs. {In addition, our results confirm that this method of constraining $\Omega_k$ is strongly dependent on lens models. However, obtaining a more accurate phenomenological model for lens galaxies is highly predictable as future massive surveys observe more and more SGL samples, which will significantly improve the constraint of cosmic curvature.}

\end{abstract}



\section{Introduction} 
The question of whether the spacial geometry of our universe being open, flat, or closed, characterized by spatial curvature parameter $\Omega_k$ corresponding to $\Omega_k>0$, $\Omega_k=0$, and $\Omega_k<0$, respectively, is a fundamental issue related to the origin and evolution of the universe. The inflationary cosmology predicts a flat universe, and this has been confirmed by the precise measurements of the cosmic microwave background (CMB) \citep{Guth:1980zm,Linde:1981mu,Bennett:1996ce}. The latest Planck 2018 results reported a very stringent constraint on the curvature parameter, $\Omega_k=0.001\pm0.002$, which is from the combination of CMB power spectra data and baryon acoustic oscillation (BAO) measurements in the framework of the $\Lambda$ cold dark matter ($\Lambda$CDM) model \citep{Planck:2018vyg}.

Although there is a precise constraint on $\Omega_k$ indicating a flat universe, two points should be noticed. First, this tight constraint depends on a specific cosmological model and is based on the early-universe measurements. The Hubble tension problem \citep{Riess:2019cxk,DiValentino:2021izs,Vagnozzi:2019ezj,Zhang:2019ylr,Qi:2019zdk,Vattis:2019efj,Zhang:2014ifa,Guo:2018ans,Zhao:2017urm,Guo:2017qjt,Guo:2019dui,Feng:2019jqa}, the most serious crisis in modern cosmology, implies the disagreements between the early universe and the late universe within the framework of modern cosmological theory \citep{Verde:2019ivm,Riess:2019cxk,DiValentino:2021izs}. Therefore, it is necessary to remeasure the curvature parameter using the late-universe observations and preferably cosmological model-independent methods. Second, recent studies \citep{DiValentino:2019qzk,Handley:2019tkm} concerning the curvature parameter found that the Planck power spectra prefer a closed universe at more than 99\% confidence level. However, combining the Planck data with BAO data prefers a flat universe, with a small error of 0.002. Conclusions regarding $\Omega_k$ from the combination of these data sets should be treated with suspicion. Thus, this further urges us to re-examine the constraints on $\Omega_k$ through a cosmological model-independent method and using low-redshift observations.

Based on the distance sum rule, \citet{Rasanen:2014mca} presented a cosmological model-independent method to constrain the cosmic curvature parameter with the combination of strong gravitational lensing (SGL) observations and Type Ia supernovae (SN Ia) data and obtained a $\Omega_k$ value close to zero but with poor precision. Subsequently, this method has been fully implemented with larger SGL and SN Ia samples \citep{Liu:2020bzc,Xia:2016dgk,Li:2018hyr,Wang:2019yob,Zhou:2019vou} as well as other distance indicators such as intermediate luminosity quasars \citep{Qi:2018aio}. However, the results of these previous works on the constraints of $\Omega_k$ are not consistent. For instance, with a prior from CMB observations, $\Omega_k \leq -0.1$, \citet{Rasanen:2014mca} and \citet{Xia:2016dgk} obtained that $\Omega_k$ is close to zero. However, without the prior from CMB, \citet{Li:2018hyr} constrained $\Omega_k$ with a larger SN Ia sample and found that a closed universe is preferred. The reason for this inconsistency is probably the addition of the CMB prior. Alternatively, the bias of estimation for $\Omega_k$ could also be caused by the limited number of available SGL samples bringing unknown systematic errors. Specifically, to constrain $\Omega_k$ using the distance sum rule requires calibrating the distances of lenses and sources in SGL systems by using other distance indicators. The maximum redshift of the distance indicators determines the number of SGL systems that can be calibrated. At present, the maximum redshift of sources in the observed SGL sample is about 3.6, while the maximum redshift of the SN Ia sample used commonly as a distance indicator is only about 2.3, which means that some SGL systems cannot be calibrated. Therefore, we need other distance probes capable of detecting higher redshifts. On the other hand, a disadvantage of SN Ia is that it cannot provide absolute distance unless calibrated by the distance ladder. Therefore, it is necessary to develop other reliable cosmological probes to constrain $\Omega_k$.

The successful detections of gravitational waves (GWs) \citep{LIGOScientific:2016aoc,LIGOScientific:2017vwq} bring us into the era of GW astronomy and multi-message astronomy. The absolute luminosity distance can be determined by analysing GW's waveform, which is referred to as standard siren \citep{Schutz:1986gp}. For a comparison, for SN Ia, only relative distances can be obtained. If the redshift of GW event is obtained through the electromagnetic (EM) counterpart or its host galaxy, the distance-redshift relation can be established, which is of importance for cosmological studies \citep{Qi:2019spg,Qi:2019wwb,Zhao:2010sz,Wang:2018lun,Zhang:2019ylr,Wang:2019tto,Zhang:2019loq,Zhang:2019ple,Zhao:2019gyk,Jin:2022tdf,Jin:2022qnj,Jin:2020hmc,Wang:2021srv,Jin:2021pcv,Bian:2021ini}. According to the conservative estimates, the third-generation ground-based GW observatory, such as the Einstein Telescope (ET) with one order of magnitude more sensitive than the current GW detectors, can detect 1000 GW events with the redshift information from the binary neutron star (BNS) mergers in a ten-year observation \citep{Nissanke:2009kt,Zhao:2010sz,Cai:2016sby,Zhao:2017cbb,Chen:2020zoq}. Moreover, the detectable redshifts of GWs could reach much higher. It is no doubt that the observations of GWs will become an important tool for cosmological studies in the near future.

Considering the above model-independent constraints on $\Omega_k$ based on the method of distance sum rule, GW observation could provide a perfect complement to traditional cosmological probes. Therefore, in this paper, we will investigate how GWs as a distance indicator will affect the constraints on $\Omega_k$ in the near future of GW astronomy. Our investigation includes two parts. First, based on ET in its 10-year observation, we simulate 1000 GW standard sirens and constrain $\Omega_k$ in combination with the latest observed SGL sample. {Since this method is dependent strongly on the lens models characterizing the mass distribution of lens galaxies \citep{Qi:2018aio}, we will perform the constraint on $\Omega_k$ in three lens models extensively used in strong lensing studies.} Next, we consider the possible developments of next decades. During the construction and subsequent observation of ET, the ongoing and future massive surveys like Large Synoptic Survey Telescope (LSST) or Dark Energy Survey will provide a large sample of well-measured SGL systems. For example, according to the prediction of~\citet{Collett:2015roa}, the LSST survey could potentially observe $1.2\times10^5$ SGL systems. In this paper, we also make a forecast for what constraints on $\Omega_k$ can be obtained with such a significant increase of the number of SGL systems.

\section{METHODS AND DATA}

\subsection{Distance sum rule}

According to the cosmological principle that the universe is homogeneous and isotropic at large scales, the spacetime geometry can be described by the Friedmann-Lema\^\i tre-Robertson-Walker (FLRW) metric, so we have
\begin{equation}\label{RW}
ds^{2}=-dt^{2}+a^{2}(t)\left\{\frac{dr^{2}}{1-kr^{2}}+r^{2}(d\theta^{2}+\sin^{2}\theta d\phi^2)\right\},
\end{equation}
where $a(t)$ denotes the cosmic scale factor, and $k$ is a constant associated with the spatial curvature. Considering a SGL system in the FLRW metric,  the angular diameter distance between the lens galaxy at redshift $z_{l}$ and the source at redshift $z_{s}$ can be represented as $D_{A}(z_{l}, z_{s})$. The dimensionless comoving distance $d(z)$ between the lens and the source can be described as 
 \begin{equation}\label{dz_DA}
  \begin{aligned}
  d(z_{l}, z_{s})&=(1+z_{s})H_{0}D_{A}(z_{l}, z_{s})\\
  &=\frac{1}{\sqrt{|\Omega_{k}|}} {\rm{sinn}} \left[\sqrt{|\Omega_{k}|} \int_{z_{l}}^{z_{s}}\frac{H_{0 }dz^{\prime}}{H(z^{\prime})}\right],\\
  &
  \end{aligned}
  \end{equation}
where
  \begin{equation}
  {\rm sinn}(x)=
  \begin{cases}
  \sin(x), & \text{$\Omega_{k}<0$},\\
  x, & \text{$\Omega_{k}=0$},\\
  \sinh(x), & \text{$\Omega_{k}>0$}.
  \end{cases}
  \end{equation}
Here, $H(z)$ is the Hubble parameter, and $H_{0}$ is the Hubble constant. $\Omega_{k}={-k}/({H_{0}^{2}}a_{0}^{2})$ ($a_{0}=a(0)$) is the spatial curvature parameter. For convenience, we define $d_{l}=d(0, z_{l})$, $d_{s}=d(0, z_{s})$, and $d_{ls}=d(z_{l}, z_{s})$. These three dimensionless distances in the FLRW universe and cosmic curvature $\Omega_k$ satisfy the distance sum rule \citep{Bernstein:2005en,Rasanen:2014mca}: 
\begin{equation}\label{dls_dl}
\frac{ d_{ls}}{ d_{s}}=\sqrt{1+\Omega_{k}d_{l}^{2}}-\frac{d_{l}}{d_{s}}\sqrt{1+\Omega_{k}d_{s}^{2}}.
\end{equation} 
Obviously, we obtain $d_{s} = d_{l} + d_{ls}$ if the universe is spatially flat ($\Omega_{k}=0$). Simultaneously, $d_{s} < d_{l} + d_{ls}$ and $d_{s} > d_{l} + d_{ls}$ correspond to a spatially closed ($\Omega_{k}<0$) and open ($\Omega_{k}>0$) universe, respectively. On the basis of Equation~(\ref{dls_dl}), if we obtain the distances $d_{l}$, $d_{s}$, and $d_{ls}$ from observations, the spatial curvature $\Omega_{k}$ can be directly derived without any assumption regarding the specific cosmological model. In this work, the distances $d_{l}$ and $d_{s}$ are inferred from the GW data, while the distance ratio $d_{ls}/d_{s}$ can be obtained from the observations of SGL systems. 

\subsection{Data simulation for gravitational wave standard sirens}

All the GW events considered in this work are assumed to be produced by the mergers of binary neutron stars (BNSs). The neutron star (NS) mass distribution is randomly sampled in the interval [1, 2] $M_{\odot}$, where $M_{\odot}$ is the solar mass, the same as in the literature \citep{Cai:2017aea,Zhang:2018byx,Wang:2018lun}. The redshift distribution of GW sources takes the form \citep{Zhao:2010sz, Cai:2016sby}
\begin{equation}\label{p}
P(z) \propto \frac{4 \pi d_{C}^2(z) R(z)}{H(z)(1+z)}, 
\end{equation}
where $d_{C}(z)$ represents the comoving distance at the redshift $z$. $R(z)$ indicates the time evolution of the burst rate, which is given by \citep{Schneider:2000sg, Cutler:2009qv} 
\begin{equation}\label{R}
R(z)=  \begin{cases}
  1+2z, & \text{$z \le 1$},\\
  \frac{3}{4}(5-z), & \text{$1 < z < 5$},\\
  0, & \text{$z \ge 5$}.
  \end{cases}
\end{equation}

After knowing the redshift and mass distributions described above, we can generate the mock catalog of the GW standard sirens. The luminosity distance $D_{L}$ can be extracted from the GW amplitude, and its value in this simulation can be obtained by
\begin{equation}\label{ET_DL}
D_{L}(z)=(1+z)\int_{0}^{z}\frac{dz^{\prime}}{H(z^{\prime})}.
\end{equation}
In this simulation, the fiducial cosmological model we choose is the flat $\Lambda$CDM universe and the values of parameters are taken from Planck 2018 results \citep{Planck:2018vyg}.

For the estimation of the luminosity distance error $\Delta{D_{L}}$, it depends on the sensitivity of the GW detector and the signal-to-noise ratio (SNR) of a GW event. The strain $h(t)$ in GW interferometers quantifies the difference of two optical paths due to the passing of GW, following \cite{Sathyaprakash:2009xs} and \cite{Zhao:2010sz}, which can be denoted as
  \begin{equation}\label{interferometers}
  h(t)=F_{+}(\theta, \phi, \psi)h_{+}(t)+F_{\times}(\theta, \phi, \psi)h_{\times}(t),
  \end{equation}
where $\psi$ is the polarization angle, and $(\theta,\phi)$ describe the source-location angles relative to the detector. Here, the antenna pattern functions $F_{+}$ and $F_{\times}$ of the ET are written as \citep{Cai:2016sby}
\begin{equation}\label{F}
\begin{aligned}
  F_{+}^{(1)}(\theta, \phi, \psi)=&\frac{\sqrt{3}}{2} \bigg[\frac{1}{2}(1+\cos^{2}(\theta))\cos(2\phi)\cos(2\psi)\\&- \cos(\theta)\sin(2\phi)\sin(2\psi)\bigg],\\
  F_{\times}^{(1)}(\theta, \phi, \psi)=&\frac{\sqrt{3}}{2} \bigg[\frac{1}{2}(1+\cos^2(\theta))\cos(2\phi)\sin(2\psi)\\&+ \cos(\theta)\sin(2\phi)\cos(2\psi)\bigg].
\end{aligned}
\end{equation}
There are three interferometers with $60^{\circ}$ inclined angles for each other, with $F_{+, \times}^{(2)}(\theta, \phi, \psi)=F_{+, \times}^{(1)}(\theta, \phi+\frac{2\pi}{3}, \psi)$ and $F_{+, \times}^{(3)}(\theta, \phi, \psi)=F_{+, \times}^{(1)}(\theta, \phi+\frac{4\pi}{3}, \psi)$.

Then, the Fourier transform $\mathcal H(f)$ of the time domain waveform $h(t)$ can be derived as \citep{Zhao:2010sz}
\begin{equation}\label{H}
   \mathcal H(f)= \mathcal Af^{-7/6} \exp[i(2\pi ft_{0}-\pi/4 +2\Psi(f/2)-\varphi_{(2.0)})].
\end{equation}   
Here, the definitions of the funtions $\Psi$  and $\varphi_{(2.0)}$ can be found in \cite{Zhao:2010sz}. The Fourier amplitude $\mathcal A$ is defined as
\begin{equation}\label{A}
\begin{aligned}
  \mathcal A= &\frac{1}{D_{L}}\sqrt{F_{+}^2(1+\cos^{2}(\iota))^2 +4F_{\times}^2\cos^2(\iota)} \\&\times \sqrt{5\pi/96}\pi^{-7/6}\mathcal M_{c}^{5/6},
\end{aligned}
\end{equation}
where $\mathcal M_{c}=(1+z)M\eta^{3/5}$ is chirp mass. Here, $M$ is the total mass of the coalescing binary with component masses $m_{1}$ and $m_{2}$, namely $M=m_{1}+m_{2}$, and $\eta=m_1m_2/(m_1+m_2)^2$. The parameter $\iota$ is the inclination angle of the binary's orbital angular momentum with the line of sight, which can be obtained from the accompanying EM counterpart of the GW event like short gamma-ray bursts (SGRBs). SGRBs are believed to be strongly beamed phenomena \citep{Nakar:2005bs, Fermi-LAT:2009owx, Rezzolla:2011da}. Once SGRBs are observed, it means that the binaries should be aligned nearly face on (i.e., $\iota \simeq0$). We take the maximal inclination to be $\iota=20^\circ$. In general, one would need to compute all the Fisher matrices with random inclination angles and then select the sources above the detection threshold that happen to have an EM counterpart. However, according to the analysis of \cite{li2015extracting}, averaging the Fisher matrix over the inclination $\iota$ and the polarisation $\psi$ with the constraint $\iota \leq 20^\circ$ is approximately equivalent to taking $\iota=0$. Moreover, in the previous simulation of GW \citep{Zhao:2010sz,Cai:2017aea,Zhang:2018byx,Wang:2018lun}, the inclination $\iota$ was also treated in the same way. Following them, therefore, we set $\iota=0$ in the simulation of GW data.

After knowing a waveform of GW, one can calculate its signal-to-noise ratio (SNR). For the ET detector, a GW event is confirmed only when the SNR reaches at least 8. The combined SNR of the network including three equivalent independent interferometers can be written as
\begin{equation}\label{rho}
\rho=\sqrt{\sum_{i=1}^{3}(\rho^{(i)})^2},
\end{equation}
where $\rho^{(i)}=\sqrt{\left\langle \mathcal{H}^{(i)}, \mathcal{H}^{(i)} \right\rangle}$, and the inner product is denoted as
\begin{equation}\label{H2}
\langle a, b \rangle=4 \int_{f_{\rm lower}}^{f_{\rm upper}} \frac{\tilde{a}(f)\tilde{b}^{*}(f)+\tilde{a}^{*}(f)\tilde{b}(f)}{2} \frac{df}{S_{h}(f)},
\end{equation}\label{H3}
where a tilde represents the Fourier transform of the function. Here, $S_{h}(f)$ is the one-side noise power spectral density, and its form for ET is taken to be the same as in \cite{Freise:2009nz,Zhao:2010sz,Cai:2017aea}. For the detection rate of GW from BNS mergers with redshift measurements enabled by EM counterparts, according to the recent studies \citep{Yu:2021nvx,Chen:2020zoq} by investigating various models of the short $\gamma$-ray bursts and afterglows, a rough estimation of about 1000 GW standard sirens for the 10-year observation of ET is achievable. Although the approximation of $\iota=0$ we take above could increase the SNR, it does not increase our estimated detection rate, which is based on more robust studies about the short $\gamma$-ray bursts. Therefore, we simulate 1000 GW standard sirens based on a 10-year observation of ET.

Applying the Fisher information matrix, the instrument error of $D_{L}$ could be given by
\begin{equation}\label{Fisher}
 \Delta D_{L}^{\rm inst}\simeq \sqrt{\left \langle \frac{\partial \mathcal H}{\partial D_{L}},\frac{\partial \mathcal H}{\partial D_{L}} \right \rangle^{-1}}. 
\end{equation}
Due to $\mathcal H \varpropto D_{L}^{-1}$ as shown in Equations (\ref{H}) and~(\ref{A}), we have 
\begin{equation} \label{Dh}
\frac{\partial \mathcal H}{\partial D_{L}}=-\frac{\mathcal H}{ D_{L}}.
\end{equation} 
By substituting Equation (\ref{Dh}) into Equation (\ref{Fisher}), we can obtain
\begin{equation}
\Delta D_{L}^{\rm inst}\simeq \sqrt{\frac{D_L^2}{\left \langle \mathcal H, \mathcal H \right \rangle}} \simeq \frac{D_{L}}{\rho}.
\end{equation}
Note that the uncertainty of the inclination $\iota$ would affect the SNR, and the maximal effect of the inclination on the SNR is a factor of 2 ($0^\circ <\iota < 90^\circ$). Then, the instrumental error on the luminosity distance can be written as
\begin{equation}\label{sigma2}
\Delta D_{L}^{\rm inst}\simeq \frac{2D_{L}}{\rho}.
\end{equation}
Besides, the error from the weak lensing should be taken into account as well, wherein $\Delta D_{L}^{\rm lens}= 0.05zD_{L}$ \citep{Sathyaprakash:2009xt}.
Finally, the total error of $D_{L}$ can be expressed as
\begin{equation}\label{total sigma}
\begin{aligned}
  \Delta D_{L}&=\sqrt{(\Delta D_{L}^{\rm inst})^2+(\Delta D_{L}^{\rm lens})^2}\\
            & =\sqrt{\left(\frac{2D_{L}}{\rho}\right)^2+(0.05zD_{L})^2}.
\end{aligned}
\end{equation} 
In this way, we generate a catalogue of GW standard sirens with the redshift $z$, the luminosity distance $D_{L}$, and the error of the luminosity distance $\Delta D_{L}$.

\subsection{Gaussian process}
Using the distance sum rule to constrain $\Omega_k$ requires the knowledge of the distances in SGL systems, which usually could be implemented by the distance calibration using other distance indicators, such as GWs, as done in this paper. However, one key difficulty is that there is no one-to-one correspondence between the redshifts of SGL data and GW data. In the previous works, there are two effective ways to do this, the polynomial fitting and Gaussian process (GP). In this paper, we adopt the GP method based on GaPP Python code \citep{Seikel:2012uu, Seikel:2012cs} to reconstruct a smooth distance-redshift curve of $D_L$ from GWs so that we can calibrate the distances in SGL data.

\begin{figure}
\includegraphics[width=0.5\textwidth]{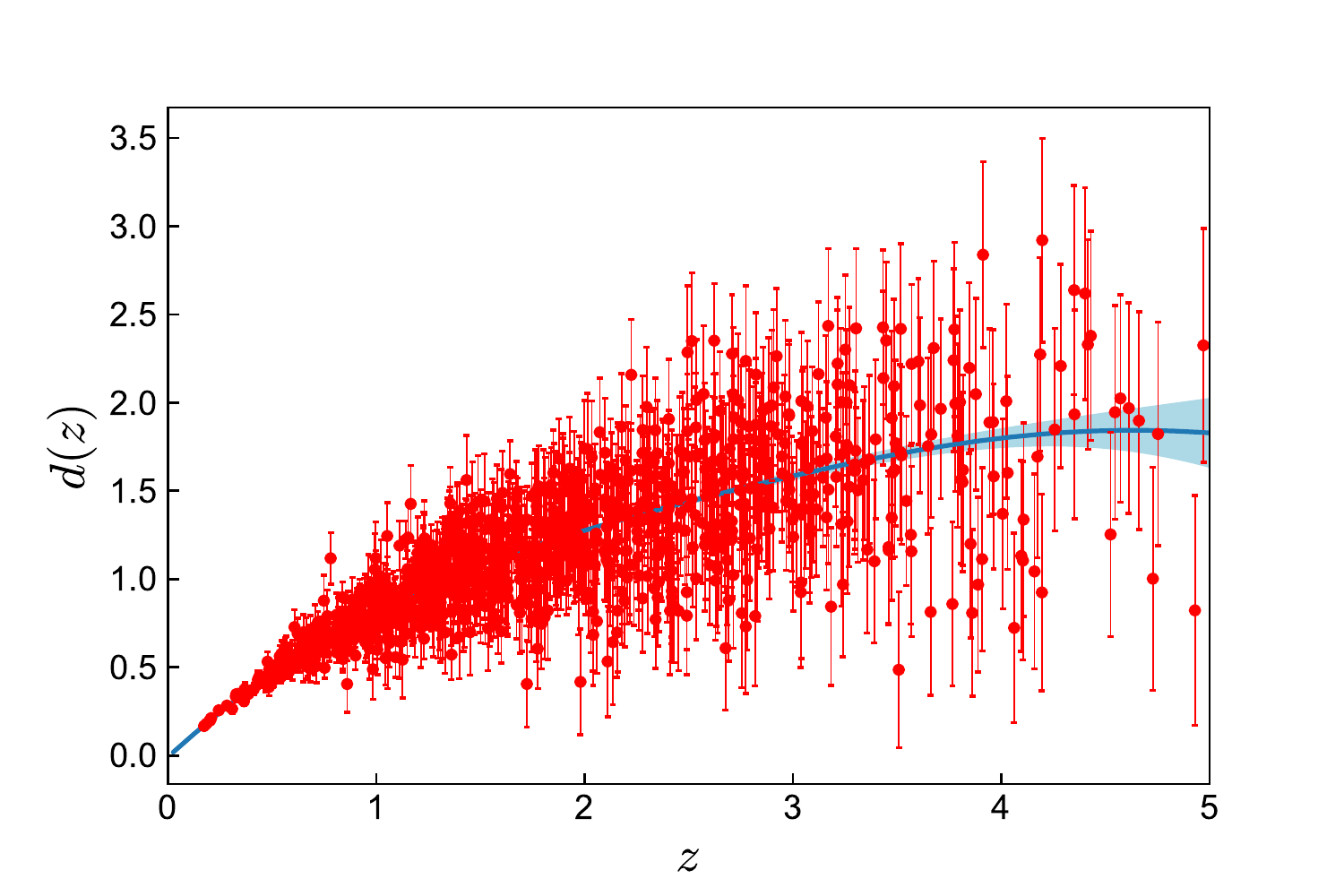}
\caption{Reconstruction of the dimensionless comoving distances from the 1000 simulated GW standard siren data. The red points with error bars represent the simulated data. The blue shaded area and the blue solid line denote the 1$\sigma$ confidence level errors and best-fit values of the reconstruction by using the GP method.
}
\label{fig:gapp_ET}
\end{figure}

This reconstruction method has been widely used in cosmology \citep{Seikel:2012uu, Seikel:2012cs,Zhang:2018gjb,Zhang:2016tto,Cai:2019bdh,Wang:2020dbt,Seikel:2013fda}, by which the reconstructed function $f(z)$ is a Gaussian distribution at each point $z$, and its values at different points $z$ and $\tilde{z}$ are connected by a covariance function $k(z,\tilde{z})$. There are various forms for the covariance function. According to the analysis in \cite{Seikel:2013fda}, the squared exponential form with the Mat\'{e}rn $(\nu=9/2)$ covariance function can lead to more reliable results than all others. So we take it here and its expression is
\begin{eqnarray}
k(z,\tilde{z})&=&\sigma_f^2\exp(-\frac{3|z-\tilde{z}|}{\ell})\nonumber\\
&\times &(1+\frac{3|z-\tilde{z}|}{\ell}+\frac{27(z-\tilde{z})^2}{7\ell^2}\nonumber\\
&+&\frac{18|z-\tilde{z}|^3}{7\ell^3}+\frac{27(z-\tilde{z})^4}{35\ell^4}),\label{7}
\end{eqnarray}
where $\sigma_f$ and $\ell$ are hyperparameters which can be optimized by the GP itself via the observational data. To determine the dimensionless distances, $d_l$ and $d_s$, in Equation (\ref{dls_dl}), firstly we convert the luminosity distances of GWs into the dimensionless distances via the following relation
\begin{equation}
d(z)=\frac{H_0D_L(z)}{(1+z)}.
\end{equation}
With the simulated GW data with redshift measurements enabled by EM counterparts, we can use the smoothing technique of GP to reconstruct the distance-redshift curve with 1$\sigma$ confidence region as shown in Figure \ref{fig:gapp_ET}. In this way, the dimensionless comoving distances corresponding to the source and lens for an SGL system could be determined by the reconstructed distance-redshift curve, as well as the errors of distances.

\subsection{Strong gravitational lensing systems}
In this subsection, we briefly introduce the SGL system and the observational SGL sample we used. For SGL systems, the measurements of lens velocity dispersion $\sigma$ could be used commonly as a statistical quantity to constrain cosmological parameters and density profiles of lens galaxies. In general, early-type galaxies are more massive and dominant in most SGL samples. Moreover, they also could be characterized by a general mass model because most of them satisfy the spherical symmetry distribution \citep{Chen:2018jcf}. With strict criteria to ensure the validity of the assumption of spherical symmetry on the lens galaxies, \citet{Chen:2018jcf} compiled a sample of SGL including 161 galaxy-scale strong lensing systems from the following surveys: the Sloan Lens ACS (SLACS) survey \citep{Bolton:2005nf,Bolton:2008xf,Auger:2009hj,Auger:2010va,Shu:2014aba,Shu:2017yon}, the Baryon Oscillation Spectroscopic Survey (BOSS) Emission-Line Lens Survey (BELLS) \citep{Brownstein:2011leg}, the BELLS for GALaxy-Ly $\gamma$ EmitteR sYstemsGALLERY \citep{Shu_2016a, Shu_2016b}. In this SGL sample, 130 SGL systems have the measurements of the luminosity density slope $\delta$ for lens galaxies which is obtained by fitting the two-dimensional power-law luminosity profile to the high-resolution imaging data from the Hubble Space Telescope. \citet{Chen:2018jcf} found that treating $\delta$ as an observable for individual lens galaxy rather than treating it as a universal parameter for all lens galaxies is necessary to get an unbiased cosmological estimate. Therefore, in this paper, we also use this truncated SGL sample including 130 SGL systems with the measurements of $\delta$, for which the redshift range of lenses is $0.0624\leq z_l\leq0.7224$ and the redshift range of sources is $0.1970\leq z_s\leq 2.8324$.

As mentioned above, the velocity dispersion of intervening galaxies is the statistical quantity for cosmological fitting, and its measurement could be obtained from the spectroscopic data. To eliminate the effect of the aperture size on measurements of velocity dispersions, $\sigma_{\rm ap}$ measured within a circular aperture with the angular radius $\theta_{\rm ap}$ should be normalized to a typical physical aperture within a circular aperture of radius $R_{\rm eff}/2$ (the half-light radius of the lens galaxy), according to the aperture correction formula \citep{Jorgensen:1995zz}, 
\begin{equation}\label{sigma0}
     \sigma_{0}=\sigma_{\rm ap}\left(\frac{\theta_{\rm eff}}{2\theta_{\rm ap}}\right)^{\xi},
   \end{equation} 
where $\theta_{\rm eff}=R_{\rm eff}/D_A(z_l)$, and $\xi$ is adopted as $\xi = -0.066 \pm 0.035$ \citep{Cappellari:2005ux}. It should be noted that the uncertainty of $\xi$ is going to feed into the total error of $\sigma_0$. In addition, considering the extra mass contribution from matters along the line of sight and the fractional uncertainty of the Einstein radius, 5\% uncertainty of velocity dispersion will be taken as the systematic error \citep{Wang:2019yob}.

For a SGL system, the gravitational mass $M_{\rm grl}^{\rm E}$ should equal to the dynamical mass $M_{\rm dyn}^{\rm E}$ within the Einstein radius $\theta_{\rm E}$. If the lens model and the cosmological distances are determined, $M_{\rm dyn}^{\rm E}$ could be inferred from the velocity dispersion, and $M_{\rm grl}^{\rm E}$ can also be inferred from the measurement of the Einstein radius. {As mentioned above, although this constraint of $\Omega_k$ is independent of cosmological models, it strongly depends on the lens models. Therefore, we will consider three lens models widely used in strong lensing studies for full analysis.}

\begin{itemize}
\item Singular isothermal sphere (SIS) model

For the simplest SIS model, the velocity dispersion can be expressed as \citep{Cao:2015qja}
\begin{equation}\label{SGL1}
\sigma_{0}^{\rm SIS} =\sqrt{\frac{\theta_{\rm E}}{4\pi f_{\rm E}^2}\frac{d_{s}}{d_{ls}}},
\end{equation}
where $f_{\rm E}$ is a phenomenological coefficient, which reflects the uncertainty due to the difference between the observed stellar velocity dispersion and the underlying dark matter, and other systematic effects. In terms of standard SIS model, the coefficient $f_{\rm E}$ is strictly equal to 1. In this paper, $f_{\rm E}$ is treated as a free parameter and it takes the range $0.8<f_{\rm E}^{2}<1.2$ according to some observations~\citep{Kochanek:1999rj,Ofek:2003sp}.

\item Extended power-law (EPL) lens model

Considering a more complex mass model, we assume that the luminosity density profile $\upsilon(r)$ differs from the total-mass density profile $\rho(r)$, and they take the forms \citep{Cao:2015qja}
\begin{equation}\label{rho_upsilon}
\rho(r)=\rho_{0}\left(\frac{r}{r_{0}}\right)^{-\gamma},~~~  
\upsilon(r)=\upsilon_{0}\left(\frac{r}{r_{0}}\right)^{-\delta},  
\end{equation}
where $r$ is the spherical radius from the lens galaxy center, $\gamma$ is the power law index of the total mass density profile treated as a free parameter, and $\delta$ is the power law index of the luminosity density profile, which has been measured for each lens in SGL sample we used in this paper. In addition, we also consider the anisotropy of the stellar velocity dispersion $\beta(r)$, which is given by
\begin{equation}\label{beta}
\beta(r)=1-\frac{\sigma_{\theta}^{2}}{\sigma_{r}^{2}}, 
\end{equation}
where $\sigma_{\theta}$ and $\sigma_{r}$ are the tangential and radial components of the velocity dispersion, respectively. According to the constraint on $\beta$ from a well-studied sample of nearby elliptical galaxies, we will treat it as a nuisance parameter and marginalize over it with a Gaussian distribution, $\beta=0.18\pm0.13$ \citep{Schwab:2009nz}. In this lens model, the velocity dispersion can be expressed as \citep{Chen:2018jcf}
\begin{equation}\label{SGL2}
\sigma_{0}^{\rm EPL}=\sqrt{\frac{\theta_{\rm E}}{2\sqrt{\pi}}\frac{d_{s}}{d_{ls}}\frac{3-\delta}{(\xi-2\beta)(3-\xi)}\left(\frac{\theta_{\rm eff}}{2\theta_{\rm E}}\right)^{2-\gamma}\left[\frac{\rm\lambda(\xi)-\beta \rm \lambda(\xi+2)}{\rm \lambda(\gamma)\rm \lambda(\delta)}\right]},
\end{equation}
where $\xi= \gamma + \delta-2$, and $\lambda(x)=\Gamma\left(\frac{x-1}{2}\right)/\Gamma\left(\frac{x}{2}\right)$. It is worth noting that if $\gamma=\delta=2$ and $\beta=0$, the EPL model will be reduced to the standard SIS model. According to the studies of previous works \citep{Ruff:2010rv,Bolton:2012uh,Cao:2016wor,Cui:2017idf,Holanda:2017jrj}, the dependence of total mass density slope $\gamma$ on the redshift is possible. Therefore, we consider two scenarios of $\gamma$ to further explore the issues we are interested in, i.e., 
\begin{description}
\item[(i)] EPL1: $\gamma=\gamma_{0}$,
\item[(ii)] EPL2: $\gamma=\gamma_{0}+\gamma_{1}  z_{l}$,
\end{description}
where $\gamma_0$ and $\gamma_1$ are free parameters.

The distance ratio $d_{ls}/d_s$ can be inferred from the distance sum rule, once the distances $d_l$ and $d_s$ are calibrated by GWs, in which the spatial curvature $\Omega_{k}$ is involved. Thus, the values of $\sigma_0$ in three lens models can be obtained. $\Omega_k$ can be constrained by maximizing the likelihood function $\mathcal{L} \propto e^{-\chi^{2} / 2}$. The $\chi^2$ function is defined as
\begin{equation}\label{chi}
\chi^{2}(\boldsymbol{p},\Omega_{k})=\sum^{N}_{i=1}\frac{[\sigma_0^{\rm lens}(z_{i},\boldsymbol{p},\Omega_{k})-\sigma_0^{\rm obs}(z_{i})]^2}{(\Delta\sigma^{\rm tot}_0)^2},
\end{equation}
where $N$ denotes the number of SGL data points, and $\boldsymbol{p}$ is the parameters of lens models. It should be noted that the total uncertainty $\sigma^{\rm tot}_0$ not only has the contribution from the measurements of SGL systems, but also contains the uncertainties from distance calibrations of $d_l$ and $d_s$.
 
\end{itemize}

\section{Results and Discussion}
By using the \texttt{emcee} Python module \citep{ForemanMackey:2012ig} based on the Markov Chain Monte Carlo (MCMC) method, we obtain the cosmological model-independent constraint on $\Omega_k$ in the framework of three lens models. Different from previous work~\citep{Rasanen:2014mca,Xia:2016dgk} considering a prior of $\Omega_{k}> -0.1$ from the CMB observation~\citep{Vonlanthen:2010cd, Audren:2012wb, Audren:2013nwa}, we do not take this prior because our motivation is to measure $\Omega_k$ using only the late-universe observations. Firstly, we present the constraint results from the current data sets of 130 SGL systems combined with the simulated GW data. Secondly, considering the upcoming LSST survey with a large sample of SGL as expected, we also forecast what constraint on $\Omega_k$ could be achieved.

\subsection{Results from current SGL data}
\begin{figure*}
\includegraphics[width=0.32\textwidth]{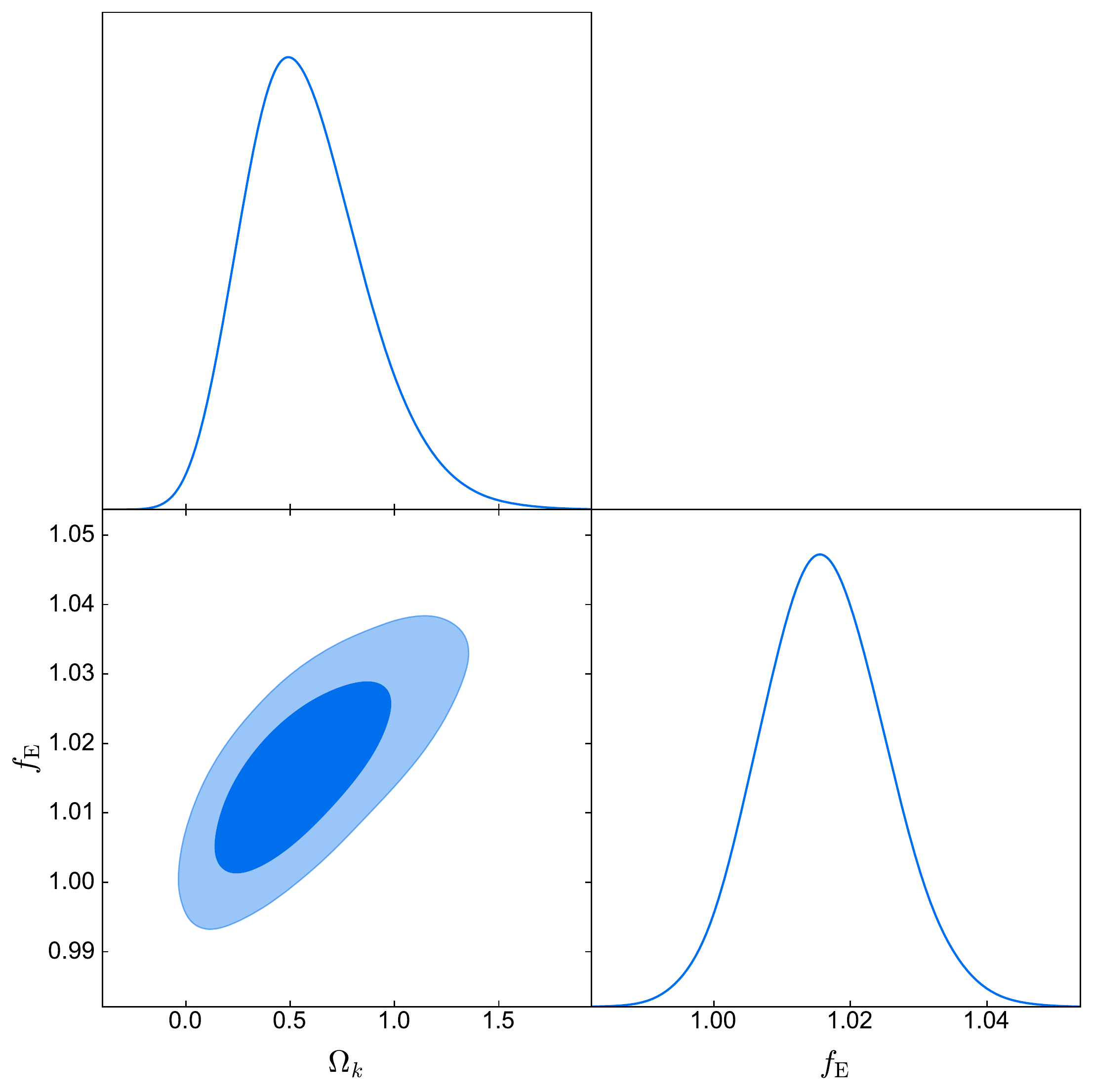}
\includegraphics[width=0.32\textwidth]{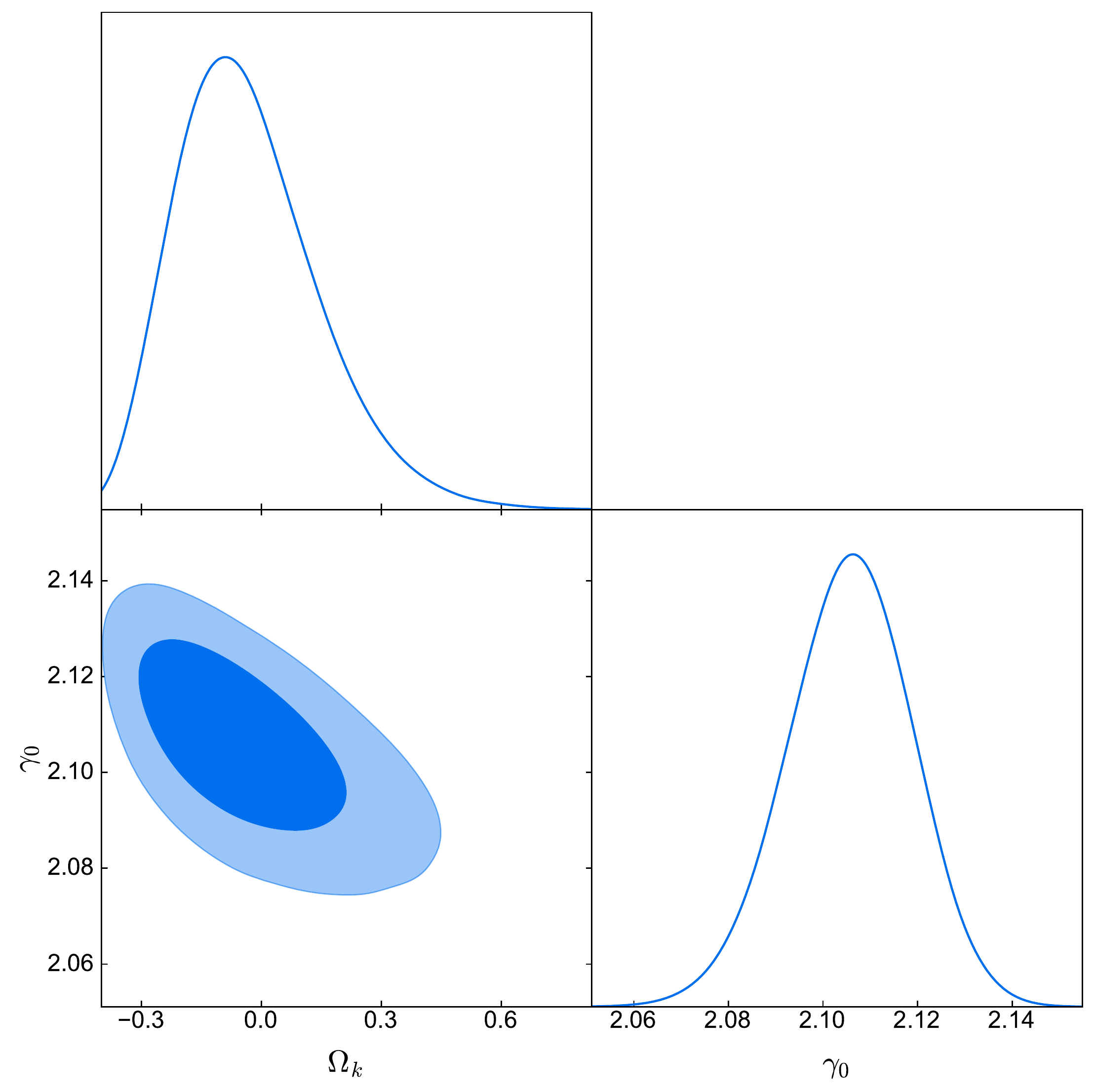}
\includegraphics[width=0.32\textwidth]{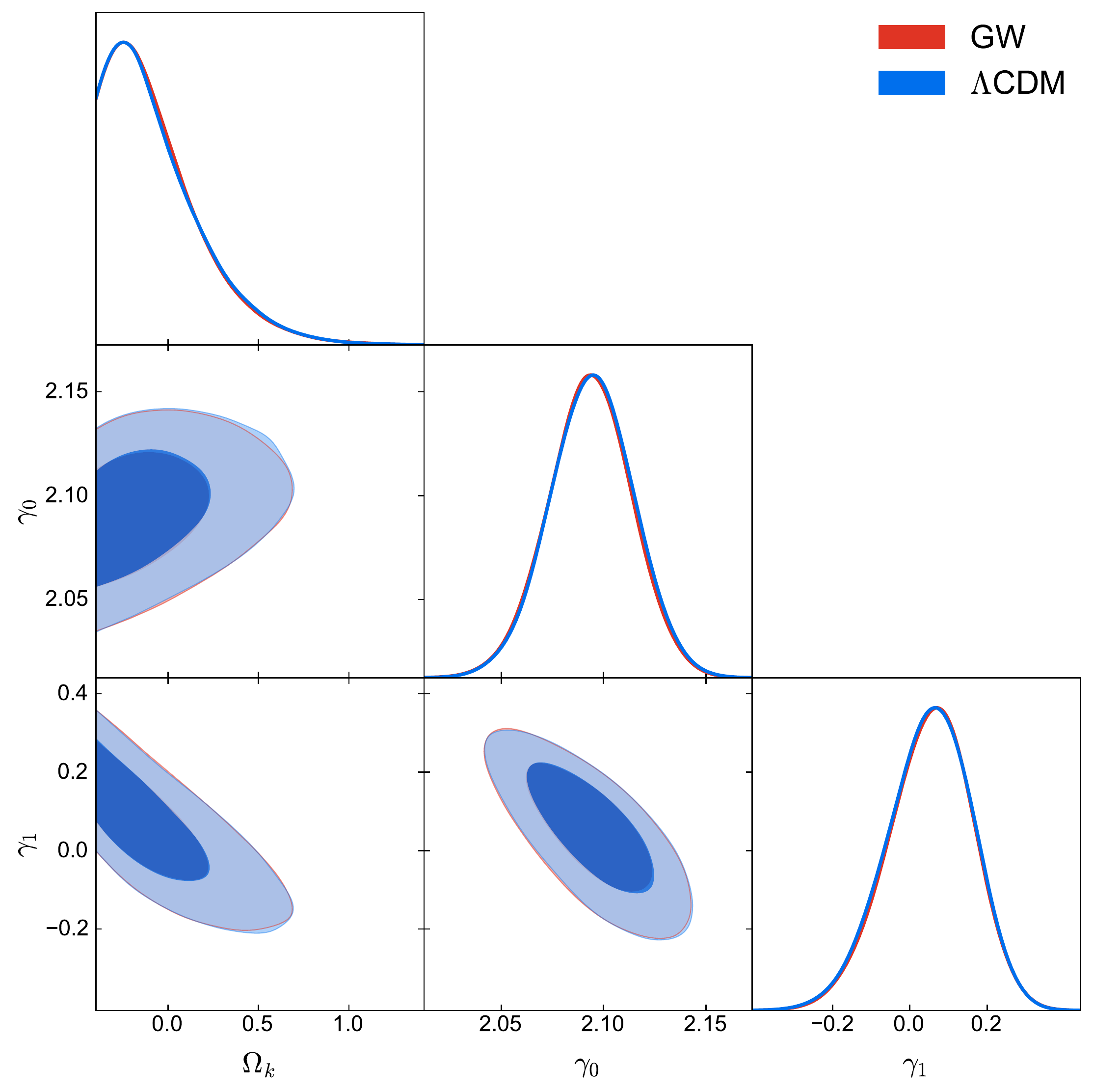}
\caption{One-dimensional and two-dimensional posterior distributions for all parameters from 130 SGL systems. 
Left: The constraints on spatial curvature $\Omega_{k}$ and the lens profile parameter $f_{\rm E}$ in the SIS lens model. 
Middle: The constraints on spatial curvature $\Omega_{k}$ and the lens profile parameter $\gamma_{0}$ in the ELP1 lens model.
Right: The constraints on spatial curvature $\Omega_{k}$ and the lens profile parameters $\gamma_{0}$ and $\gamma_{1}$ by using GW and the $\Lambda$CDM to provide the distances in the EPL2 lens model.} 
\label{fig:observation}
\end{figure*}

For the simplest SIS model, the constraints on $\Omega_k$ and $f_{\mathrm{E}}$ are shown in Figure \ref{fig:observation} and Table \ref{tab:summary}. By using the combination of 1000 GW simulation data and 130 SGL observational data, the spatial curvature parameter is constrained to be $\Omega_{k}=0.550^{+0.313}_{-0.256}$, wherein a zero value of $\Omega_{k}$ is ruled out at 2$\sigma$ confidence level. It should be noted that while the 1000 GW data are simulated in a flat universe, the 130 SGL data are actually observed, so the constraint result of $\Omega_k$ is still instructive. For the parameter $f_{\mathrm{E}}$ reflecting the mass distribution of the lens galaxies, we obtain a result of $f_{\mathrm{E}}=1.016 \pm 0.009$ at 1$\sigma$ confidence level, which is in good agreement with the standard SIS model $(f_{\mathrm{E}}=1)$ at 2$\sigma$ confidence level.

Now we focus on the constraint errors of parameters. Compared with the previous results using SN Ia as distance indicators to calibrate the distances of SGL, using GW standard sirens does not obtain competitive precision for the constraints on $\Omega_k$ in this lens model. For instance, by using the combination of 137 SGL data and Pantheon SN Ia sample, \citet{Zhou:2019vou} inferred the cosmic curvature parameter as $\Omega_{k}=0.483^{+0.239}_{-0.385}$ at 1$\sigma$ confidence level based on the SIS lens model. With 161 galactic-scale SGL systems and 1048 SN Ia data, \citet{Wang:2019yob} obtained a value of $\Omega_{k}=0.57^{+0.20}_{-0.28}$ at 1$\sigma$ confidence level in the framework of SIS lens model. Although the constraint error of $\Omega_k$ has not been significantly improved by using simulated GW data, with the increase of SGL data observed in the future, the GW standard siren observation covering a wider redshift range could calibrate more SGL systems than SN Ia, which will help reduce the statistical error for the constraint on $\Omega_k$.

\begin{table*}
\centering
\renewcommand\arraystretch{1.2}
\caption{The fit values of all parameters from 130 SGL systems at the 1$\sigma$ confidence level in the SIS, EPL1, and EPL2 models.}
\label{tab:summary}
\begin{tabular}{ccccccccccccccc}
\hline
Lens model&$\Omega_{k}$ &$f_{\rm E}$ & $\gamma_{0}$ & $\gamma_{1}$ \\ 
\hline
SIS & $0.550^{+0.313}_{-0.256}$ & $1.016 \pm0.009$ & $-$ & $-$ \\ 
EPL1 & $-0.052^{+0.194}_{-0.154}$ & $-$ & $2.106 \pm0.013$ & $-$\\
EPL2 & $-0.139^{+0.278}_{-0.172}$ & $-$ & $2.098 \pm0.019$ & $0.053^{+0.098}_{-0.108}$\\
\hline
\end{tabular}
\end{table*}

For the EPL1 model, we present the constraint results in Figure \ref{fig:observation} and Table \ref{tab:summary}. The fit value at 1$\sigma$ confidence level of $\Omega_k$ is $\Omega_{k}=-0.052^{+0.194}_{-0.154}$, in excellent agreement with a flat universe. By comparing with the results from the SIS model, we find that the model selection has a strong influence on the constraint on $\Omega_k$, which further confirms the conclusion of previous works as well \citep{Qi:2018aio,Wang:2019yob}. Moreover, for the constraint on $\Omega_k$ in the EPL1 model, we obtain a more stringent result by using GWs as the distance indicators with respect to using SN Ia. \citet{Zhou:2019vou} presented a result of $\Omega_{k}=0.100^{+0.538}_{-0.114}$, and \citet{Wang:2019yob} obtained $\Omega_{k}=0.25^{+0.23}_{-0.16}$ from the combination of 161 SGL data and 1048 SN Ia data. On the other hand, we stress that the EPL1 model will be reduced to the standard SIS model if $\gamma_0=2$. The constraint result of $\gamma_0$ we obtain is $\gamma_{0}=2.106 \pm0.013$. It is clearly shown that the SIS model has been excluded at 2$\sigma$ confidence level.

For the EPL2 model, the one-dimensional marginalized posterior distributions and the contours of parameters are shown in Figure \ref{fig:observation}, and the constraint results are summarized in Table \ref{tab:summary}. It can be clearly seen that the result $\Omega_{k}=-0.139^{+0.278}_{-0.172}$ is well consistent with a flat universe. Compared to the results of the EPL1 model, this constraint on $\Omega_k$ becomes weaker, possibly due to the addition of a parameter $\gamma_1$. However, this constraint on $\Omega_k$ is tighter than that of the SIS model, even though the number of parameters here is one more than the SIS model. All these results indicate that reasonably modeling the mass distribution of lens galaxies is an important factor for constraining $\Omega_k$ with this method. For the lens model parameters, we have $\gamma_{0}=2.098 \pm0.019$, and $\gamma_{1}=0.053^{+0.098}_{-0.108}$, wherein a zero value of $\gamma_{1}$ is included at 1$\sigma$ confidence level. This suggests that the dependence of the total mass density profile slope $\gamma$ on the redshift is not significant in this work, which supports the EPL2 lens model being reduced to the EPL1 model at $1\sigma$ confidence level.

{In our analyses, GW data are used as the distance indicator to calibrate the distances of source and lens in SGL data. For the constraint on $\Omega_k$, which of the two data (SGL or GW) is dominant needs to be clarified. First, for the best-fit values, by comparing with the previous results using SN Ia as distance indicators, we find that the best-fit values of $\Omega_k$ in the same lens model are very close, as discussed above. In addition, the simulation of GW to provide the distances is based on the flat ($\Omega_k=0$) $\Lambda$CDM model. Therefore, a flat universe under any lens model should be obtained if the GW data dominate the constraint on $\Omega_k$. However, we find that the best-fit values of $\Omega_k$ in three lens models are different. These two points indicate that the SGL data are dominant for the constrained best-fit values. Second, we explore which of the two data dominates the constrained uncertainties of $\Omega_k$. Taking the EPL2 model as an example, we perform the same constraint by using the $\Lambda$CDM model as same as the fiducial model in the simulation of GW to provide the distances instead of GW data. In the right panel of Figure \ref{fig:observation}, we find that the result from the $\Lambda$CDM model is almost the same as that from GW data, even though the distances provided by the $\Lambda$CDM model have no errors. All of these imply that the SGL data dominate the constraints of $\Omega_k$ in this approach.
}

\subsection{Results from LSST simulation sample}

\begin{figure}
\includegraphics[width=0.35\textwidth]{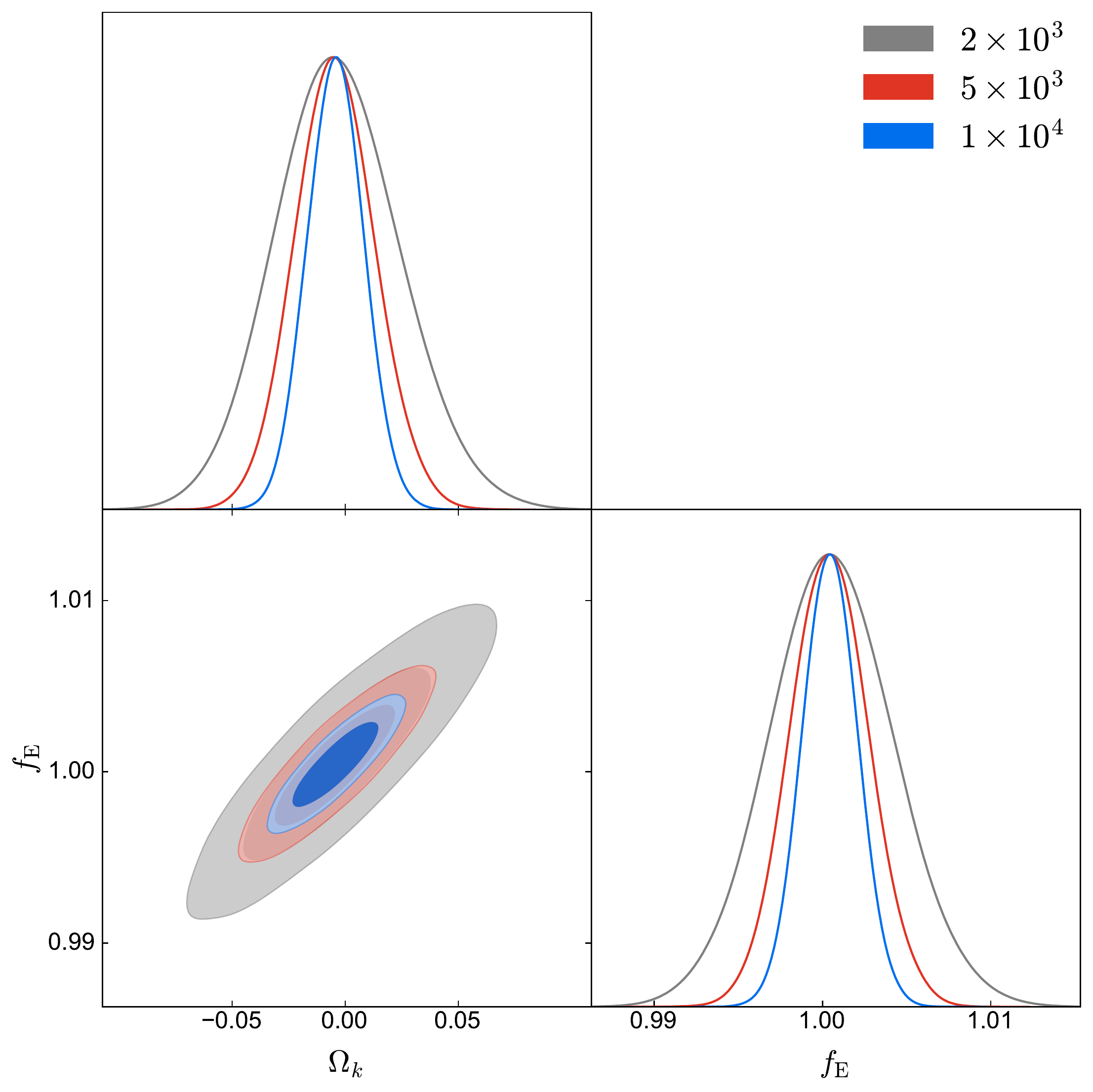}
\caption{One-dimensional and two-dimensional posterior distributions for the parameters $\Omega_{k}$ and $f_{\rm E}$ from LSST simulation samples of $2\times10^3$ (gray solid line), $5\times10^3$ (red solid line), and $1\times10^4$ (blue solid line) lenses in the SIS lens model. }
\label{fig:SIS-diffLSST}
\end{figure}
\begin{table}
\caption{\label{tab:SIS-diffLSST}The best-fit values of the parameters $\Omega_{k}$ and $f_{\rm E}$ at the 1$\sigma$ confidence level from $2\times10^3$, $5\times10^3$, and $1\times10^4$ LSST simulation SGL systems in the SIS model.}
\footnotesize\centering
\begin{tabular}{ccccccccccccccc}
\hline
Sample number & $\Omega_{k}$ & $f_{\rm E}$\\
\hline
$2 \times 10^3$ $\rm $ & $-0.004\pm 0.027$ & $1.001 \pm 0.004$\\ 
$5 \times 10^3$  $\rm $ & $-0.005 \pm 0.017$ & $1.000 \pm 0.002$\\
$1\times 10^4$  $\rm $ & $-0.004 \pm 0.012$ & $1.000 \pm 0.002$\\
\hline
\end{tabular}
\end{table}

During the construction and subsequent observations of ET, the upcoming LSST with wide field-of-view is expected to observe $1.2\times10^5$ galaxy-galaxy strong lensing. Such a large sample of SGL data is bound to produce extensive cosmological applications. Here we also make a forecast for what constraint on $\Omega_k$ can be achieved with such a tremendous increase of SGL data. Based on the performance of LSST, \citet{Collett:2015roa} performed a simulation of a realistic population of galaxy-galaxy strong lensing. For our estimations of $\Omega_k$, a fraction of the SGL sub-sample is available, considering the determination of redshift and accurate measurement on velocity dispersion, and so on. Therefore, in this paper, by using a public package LensPop\footnote{github.com/tcollett/LensPop}, we simulate $2\times10^3$, $5\times10^3$ and $1\times10^4$ well-measured SGL systems, respectively, to investigate the effect of the increase of data points in SGL sample on improving the constraints on $\Omega_k$. High-quality imaging and spectroscopic data from LSST enable highly precision inferences of Einstein radius and lens velocity dispersion. According to the analysis from \citet{Collett:2016muz}, we adopt the fractional uncertainties of observed velocity dispersion and the Einstein radius as 5\% and 3\%, respectively.

In the framework of the SIS model, the constraint results from combining GWs with $2\times10^3$, $5\times10^3$, and $1\times10^4$ mock data from LSST simulation, respectively, are shown in Figure \ref{fig:SIS-diffLSST} and Table \ref{tab:SIS-diffLSST}. We find that as the number of SGL data increases by an order of magnitude compared to the existing SGL sample, the constraint on $\Omega_k$ is improved by an order of magnitude, i.e., $\Omega_k=-0.004\pm 0.027$, from $2\times10^3$ simulated SGL systems. {This significant improvement is not only contributed by the increase of SGL samples, but also the improvement in the observation precision.} However, when the number of SGL data increases by an order of magnitude again, i.e., $\sim 1\times10^4$, the constraint on $\Omega_k$ is only improved by a factor of $\sim 2$, indicating that systematic errors will dominate over statistical errors. Although the constraint on $\Omega_k$ here is not as good as the result obtained by the combination of Planck and BAO data (with the error 0.002), it must be emphasized that our constraints are independent of any cosmological model, which will be helpful in solving cosmological tension problem concerning the cosmic curvature in the future.

\section{Conclusion} 
With the increasing precision of cosmological observations, tensions in the measurements of some key cosmological parameters has gradually emerged, which is usually viewed to be the measurement inconsistency between the early and late universe. The confusion caused by recent studies concerning cosmic curvature parameter $\Omega_k$ suggests that it is necessary to remeasure $\Omega_k$ using only the late-universe observations in a cosmological model-independent way. The distance sum rule in SGL provides such a way, provided that the distances in the sum rule can be calibrated by other observations. Usually, SN Ia can be used as a distance indicator to perform the distance calibration in this method. However, SN Ia observation has some drawbacks, such as dependence on distance ladder, narrow redshift range, and so forth. In this work, we propose that GWs can be used to provide the distance calibration in the SGL method, which can avoid the dependence on distance ladder and cover a wider redshift range. We use the simulated GW standard siren observation from the Einstein Telescope as an example to show that this scheme is feasible and advantageous.

Specifically, in the framework of three lens models, namely SIS, EPL1, and EPL2 models, we use 130 current SGL data and 1000 simulated GW standard siren data to estimate $\Omega_k$. We find that the result of the SIS model prefers an open universe at more than 2$\sigma$ confidence level, while the inferences for $\Omega_k$ in EPL1 and EPL2 models are in excellent agreement with a flat universe, which means that the lens-model selection has a strong influence on inferring $\Omega_k$. Moreover, for the constraints on $\Omega_k$ in the three lens models, we obtain the most stringent result in the EPL1 model, i.e., $\Omega_k=-0.052^{+0.194}_{-0.154}$, which is slightly tighter than that obtained by using SN Ia as distance indicators. On the whole, we find that these model-independent estimations of $\Omega_k$ using only the late-universe observations still somewhat favor a flat universe.

However, it is important to emphasize that although this constraint of $\Omega_k$ is independent on cosmological models, it is dependent strongly on lens models in fact. In this paper, the mass distribution of the lens galaxies is assumed to be spherically symmetric, which could characterize well the morphologies of early-type galaxies that are more likely to serve as intervening lenses. Although the sample of SGL we used is obtained with well-defined selection criteria to ensure the validity of the assumption of spherical symmetry, the properties of early-type galaxies as their formation and evolution are still not fully understood. There is still a long way from accurately characterizing the mass distribution of lens galaxies, which is crucial for the unbiased and precise estimation of $\Omega_k$ in this way. Fortunately, as future massive surveys observe more and more SGL samples, a more accurate phenomenological model for lens galaxies could be obtained, which will greatly improve the constraint on cosmic curvature.

Then, we further forecast what constraint can be achieved for the spatial curvature in the near future by GW standard sirens from ET and abundant SGL data from the forthcoming LSST survey. We find that about $1 \times 10^4$ SGL data combined with 1000 GW standard sirens could achieve a precise constraint of $\Delta\Omega_{k} \simeq10^{-2}$. 
Our results show that the observations of SGL and GWs by the next-generation facilities would improve the late-universe measurement of cosmic curvature by one order of magnitude.


\section*{Acknowledgements}
We would like to thank Ling-Feng Wang, Yun Chen, Shang-Jie Jin, and Dong-Ze He for helpful discussions. This work was supported by the National Natural Science Foundation of China (Grants Nos. 11975072, 11835009, and 11875102), the Liaoning Revitalization Talents Program (Grant No. XLYC1905011), the Fundamental Research Funds for the Central Universities (Grant Nos. N2005030 and N2105014), the National 111 Project of China (Grant No. B16009), and the science research grants from the China Manned Space Project (Grant No. CMS-CSST-2021-B01).

\section*{DATA AVAILABILITY}
The data underlying this article will be shared on reasonable request to the corresponding author.

\bibliographystyle{mnras}
\bibliography{bibOmegak} 


\bsp	
\label{lastpage}
\end{document}